\documentstyle[prc,preprint,aps]{revtex}
\begin{document}
\draft
\title{Estimating the nuclear level density with the 
Monte Carlo shell model}
\author{W.~E. Ormand}
\address{Department of Physics and Astronomy, 202 Nicholson Hall,\\
Louisiana State University, Baton Rouge, LA 70803-4001}
\maketitle
\begin{abstract}
A method for making realistic estimates of the density of levels in even-even 
nuclei is presented making use of the Monte Carlo shell 
model (MCSM). The procedure follows three basic steps: (1) computation of 
the thermal energy with the MCSM, (2) evaluation of the partition
function by integrating the thermal energy, and (3) evaluating 
the level density by performing the inverse Laplace transform of the
partition function using Maximum Entropy reconstruction techniques. 
It is found that results obtained with schematic interactions, which do not
have a sign problem in the MCSM, compare well with realistic shell-model 
interactions provided an important isospin dependence is accounted for.   
\end{abstract}

The density of levels in nuclei plays
an important role in understanding compound nuclear reactions. Two 
particularly important examples are the decay of the giant-dipole resonance 
(GDR) in hot nuclei~\cite{r:r1}, 
and the radiative capture of light nuclei, i.e., protons, 
neutrons, and alphas, in nucleosynthesis~\cite{r:r2}. 
In the first case, properties
of the GDR, in particular the damping width, have been studied in several
nuclei for excitation energies ranging from 50 to 200 MeV, and it has been
shown that the analysis of experimental data is very sensitive to the
the dependence of the level density on excitation energy~\cite{r:r3}.
In contrast to the GDR studies, the particle 
capture probability, which determines the rate at which nucleosynthesis 
reactions occur, is sensitive to the level density near the 
particle-decay threshold: i.e., $\sim 5-15$ MeV. 

In most applications where the level density is required, the 
Fermi-gas model estimate~\cite{r:r4} is employed
\begin{equation}
\rho(E)=\frac{\sqrt{\pi}}{12a^{1/4}E^{5/4}}\exp(2\sqrt{aE}),
\label{e:FGr}
\end{equation}
where $E$ is the excitation energy, and $a$ is the level-density parameter,
which is determined by the number of single-particle states 
at the 
Fermi energy. The principal shortcoming of the Fermi gas estimate is that
interactions between nucleons are ignored.
Effects due to  
shell corrections and pairing correlations are approximated in 
Eq.(\ref{e:FGr}) by replacing the excitation energy $E$ by backshifted 
quantity $E-\Delta$~\cite{r:r5}. 
Emprically, both $a$ and $\Delta$ exhibit a dependence on $E$ and
the number of nucleons, $A$, that cannot simply be estimated within the 
context of the Fermi-gas model; a typical value for $a$  at low excitation
energies is $a\sim A/8$.

An alternative model that explicitly includes both  one- and 
two-body correlations is the shell model. State-of-the-art 
shell-model Hamiltonians, such as the universal {\it sd}-shell (USD) 
Hamiltonian of Wildenthal~\cite{r:r6}
have been very successful at describing both excitation energies and transition
amplitudes for states in a wide range of nuclei  
($18 \le A \le 48$) up to excitation energies of the order 5-10 MeV. Although
the shell model might appear to be the obvious method for 
estimating the level density at low excitation energies, direct diagonalization
of the Hamiltonian  
faces severe computational limitations due to the fact that the number of 
basis states scales as the exponential of the number of 
valence particles. Indeed, the large number of basis states was the 
motivation behind the development of the spectral distribution methods of
French and others~\cite{r:r7}, 
which rely on the statistical properties of the elements
of the Hamiltonian matrix. A central-limit theorem can be applied to
describe the action of the Hamiltonian 
in large spaces~\cite{r:r7a}; generally reducing the problem to
the calculation of the first and second moments of the Hamiltonian.
In many practical applications, however, this limiting situation may not be 
sufficiently realized. For example, because of features of the 
Hamiltonian, which may be thought of as shell corrections, significant
departures from ``normality'' may be observed at low excitation energies.
As a consequence, it is necessary to compute higher-order moments of the 
Hamiltonain or partition the shell-model space into smaller subspaces.
Unfortunately, not only are these higher-order moments more difficult 
to evaluate, but the level density reconstructed with orthogonal Hermite 
polynomials may fail to be positive definite~\cite{r:r7b}. 
An alternative method was proposed by 
Pluha\v r and Weidenm\"uller~\cite{r:r8} in which the partial level
densities in the subspaces were assumed to have a
form predicted by the Gaussian orthogonal ensemble (GOE), i.e., semicircular,
as determined from the first and second moments of the Hamiltonian within the 
projected subspaces. The total level density, which in their method is 
guaranteed to be positive definite,
is then obtained by combining the various subpartitions with the coupling
between the subspaces being determined statisitcally from the mean 
off-diagonal matrix elements.
This procedure also faces several limitations because of the 
reliance on the GOE limit for the subspaces, as well as being 
restricted to the first and second moments of the Hamiltonian. 
Consequently, in applications to more general shell-model problems, 
this procedure tends to lead to level densities that are
somewhat broader than the exact results~\cite{r:r9}.

In this work, a method for making realistic estimates of shell-model
level densities using the Monte Carlo shell model~\cite{r:r10} (MCSM)
is presented. The power of the MCSM is that it is capable of providing
{\it exact} results for a range of observables in
model spaces where the dimensions are prohibitive for direct diagonalization.
In addition, the MCSM is quite well suited to compute thermal 
properties, such as the energy, from which the partition
function may be obtained, which then yields the level density through an  
inverse Laplace transform. The applicability of the MCSM to
the most general of shell-model Hamiltonians, however, is limited because of 
the sign problem associated with the Monte Carlo weight function. 
One is then  
faced with either using an extrapolation method~\cite{r:r11}, which tends to 
yield larger statistical
errors, or a schematic interaction that is free of the sign problem.
Here, it will be shown that schematic Hamiltonians, such as a
surface-delta interaction~\cite{r:r12}, 
possess most of the global, or collective, 
features exhibited by ``realistic'' Hamiltonians 
with one important exception: in the spectra of even-even nuclei, the higher 
isospin states tend to be too low in energy, thereby 
compressing the total level density. 
This improper isospin dependence can be corrected by  
adding the term $a\hat T^2$ to the schematic interaction,
thus shifting the excitation energy of the higher isospin states.
Unfortunately, this ${\hat T}^2$ term also has a bad sign, and
cannot be computed directly. To address this problem, a simple and 
accurate approximation for 
correcting the thermal energy for the $T^2$ dependence in even-even nuclei
is presented, and it will be shown by direct comparison 
that an MCSM calculation using a 
schematic interaction
yields a reasonable estimate of the level density obtained in 
``realistic'' shell-model calculations.

The procedure consists of four steps: (1) using a 
semi-realistic 
schematic interaction, compute the thermal expectation value of the
Hamiltonian,
\begin{equation}
E(\beta)=\frac{{\rm Tr}[\hat He^{-\beta\hat H}]}{{\rm Tr}[e^{-\beta\hat H}]}
=\frac{\sum_iE_ie^{-\beta E_i}}{\sum_ie^{-\beta E_i}}
=\frac{\int dEe^{-\beta E}E\rho(E)}{\int dEe^{-\beta E}\rho(E)},
\label{e:energy}
\end{equation}
with the MCSM; (2) correct $E(\beta)$ for the missing $T^2$ 
dependence using Eq.~(\ref{e:tcorr}) below; (3) compute the partition 
function, $Z(\beta)$, via 
\begin{equation}
\ln Z(\beta)=-\int_0^\beta d\beta^\prime E(\beta^\prime)+\ln Z(0),
\label{e:partition}
\end{equation}
where $Z(0)$ is actually the total number of states;
and (4) calculate $\rho(E)$ using maximum-entropy reconstruction techniques
to perform the inverse Laplace transform of $Z(\beta)$.

The primary goal of this work is to establish the feasibility of the 
method for making estimates of the level density in realistic situations.
To accomplish this, comparisons with exact results are necessary, and 
the focus of this work is the nucleus $^{24}$Mg, which has four
valence protons and neutrons occupying   
the $0d_{5/2}$, $1s_{1/2}$, and $0d_{3/2}$ valence 
orbitals. Because of the overall
success of the USD interaction, which was explicitly developed for this model 
space, this interaction will be used as a benchmark for success.
The schematic interaction is composed of the three single-particle 
energies and a two-body potential given by the surface-delta interaction 
(SDI)~\cite{r:r12}, which has the form
\begin{equation}
V({\bf r}_1,{\bf r}_2)=-4\pi V_0\delta({\bf r}_1-{\bf r}_2)\delta(r_1-R_0),
\end{equation}
where ${\bf r}_i$ is the position vector for the $i^{th}$ particle and
$R_0$ is the nuclear radius. The principal feature of the SDI is that it is 
basically comprised of multipole-multipole terms, with the dominant multipoles
being monopole and quadrupole. Taking 
$R_0=3.145$~fm, $V_0=54.76$~MeV/fm$^2$, 
and evaluating the two-body matrix elements
using harmonic oscillator wave functions with $\hbar\omega =13.531$~MeV, 
the $T=1$ USD matrix elements are reproduced 
by better than 500 keV. The  
single-particle energies are adjusted to reproduce the low-lying 
experimental spectra for $^{19}$O and 
the USD shell-model spectra for $^{25}$O and $^{27}$O. 
Using fixed single-particle energies across the shell and the USD mass
scaling of $(18/A)^{0.3}$ for the two-body part, the SDI Hamiltonian 
reproduces the USD spectra and level densities for oxygen isotopes
reasonably well. On the other hand, an important component  
is missing, as it is not possible to reproduce the 
binding energies for the oxygen isotopes. This feature is generic
to schematic interactions of the form multipole plus pairing, and may be 
``fixed'' by adding a term dependent on the square of the
isospin, $a{\hat T}^2$, as is illustrated in Fig.~\ref{fig:fig1}, 
where the binding 
energies for the {\it sd}-shell oxygen isotopes are compared for the USD 
interaction and the schematic Hamiltonian, with parameters 
$\epsilon_{d5/2}=-4.820$~MeV, $\epsilon_{s1/2}=-2.820$~MeV, 
$\epsilon_{d3/2}=1.530$~MeV, $V_0$=54.76~MeV/fm$^2$, and $a=0.546$~MeV. 

From the standpoint of estimating the level
density, an adequate Hamiltonian would in fact be an extension of
the SDI that includes isospin-dependent components: in particular 
the modified surface delta interaction~\cite{r:msdi}. 
Because of the sign problem in the MCSM, however, it
is generally not possible to treat isospin-dependent terms in the Hamiltonian.
On the other hand, for a given isospin value, the 
SDI spectra compare well with the realistic USD results. Hence, a reasonable 
first-order correction may be obtained by simply 
shifting the higher isospin states by adding $a\hat T^2$ to the
Hamiltonian. 
The magnitude of the coefficient $a$, however, cannot be
estimated {\it a priori}, as it seems to differ from nucleus to nucleus and 
depends on the number of the protons and neutrons as well as whether
the shell is more than half full. This is illustrated in 
Table~\ref{tab:tab1}, where the
excitation energies obtained with the SDI and USD interactions for 
the first $J=0^+$ states of each isospin are tabulated for $^{24}$Mg
ans $^{32}$S. 
Also, given in the table are the total number of states of each isospin
that contribute to the total level 
density. Because of this unpredictable $T^2$ behavior, a shell-model 
calculation with a realistic interaction, 
even within a truncated model space, for a single angular 
momentum for a few isospins is necessary to determine a 
reasonable estimate for $a$. For $^{24}$Mg, the improper isospin dependence
is corrected on average with $a=1.507$. On the other hand, to illustrate 
that the SDI interaction gives a good representation of the low-lying 
collective behavior for $^{24}$Mg,
the excitation energies of the lowest few $T=0$ states are 
compared with the USD values in Table~\ref{tab:tab2}.

Using the Monte Carlo shell model (MCSM) techniques described in 
Ref.~\cite{r:r10}, the
expectation values of observables such as the Hamiltonian, 
${\hat T}^2$, etc., were evaluated for the SDI interaction as a function of 
the inverse temperature, $\beta$, in the range $0\le \beta \le 1$~MeV$^{-1}$
in increments of $\Delta\beta=1/80$~MeV$^{-1}$. In order to ensure sufficient 
accuracy, 2000 Monte Carlo samples were taken for each $\beta$ value, and
multiple time slices were used in all MCSM calculations with 
$\beta \ge 0.0375$~MeV$^{-1}$, with the maximum number of forty being used 
at $\beta=1$~MeV$^{-1}$. Typical Monte Carlo uncertainties for the 
energy observable ranged from 10~keV for small values of $\beta$ to about
120~keV for $\beta\sim 1$~MeV$^{-1}$. 

With $E(\beta)$, the partition 
function is obtained via Eq.~(\ref{e:partition}), and 
the level density is then given by the inverse Laplace transform of 
$Z(\beta)$.
Although a saddle-point approximation may be employed, giving
$\rho(E)=e^{\beta E+\ln Z}/\sqrt{-2\pi \partial E/\partial\beta}$,
this method tends to be somewhat unstable at low excitations energies
due to difficulties associated with computing the derivative in the
denominator. An alternative method is to evaluate the inverse
transform using maximum-entropy (MaxEnt) reconstruction 
techniques~\cite{r:r13}. 
The starting point is to bin $\rho(E)$ into $N_R$ bins of equal 
width $\Delta E$, namely
\begin{equation}
\rho(E)=\sum_i^{N_R}f_i\left[\theta(E-(E_i-\Delta/2))-
\theta(E-(E_i+\Delta/2))\right],
\end{equation}
where $f_i$ is the number of levels contained within 
the $i^{th}$ bin. With this level density, the reconstructed partition 
function is then
\begin{equation}
Z_R(\beta)=\sum_i^{N_R} \frac{2f_i}{\beta}e^{-\beta E_i}
\sinh(\beta\Delta E/2).
\end{equation}
The goal of MaxEnt is to find the set of values $\{f_i\}$ that 
maximize the
extended entropy functional $\alpha S-\chi^2/2$, where 
\begin{equation}
\chi^2=\sum_j^N\frac{(Z(\beta_j)-Z_R(\beta_j))^2}{\sigma_j^2},
\end{equation}
quantifies how well the reconstruction reproduces the $N$ calculated
values of the $Z(\beta_j)$ (note $N_r\le N$) 
and the information entropy, $S$, is given by 
\begin{equation}
S=\sum_i^{N_R}(f_i-D_i-f_i\ln(f_i/D_i)).
\end{equation}
In the MaxEnt method, it is necessary to specify a default model
$\{D_i\}$, which may be used to characterize any prior information known
about the problem at hand. 
In this case, it is well known that within a finite model space, the
level density exhibits a Gaussian character~\cite{r:r14}, which may be used to
define the default model. For finite-space, shell-model calculations, 
the total number of
states is known and the first and second moments of the level density
may be obtained from $E(\beta=0)$ and 
$dE(\beta)/d\beta |_{\beta=0}$, respectively.
The reconstructed $\{f_i\}$ also depend on $\alpha$, which governs the
relative weight between the default model and chi-square. Here,
$\alpha$ was chosen so that $\chi^2\sim N$. Finally, the uncertainty in 
the $f_i$ values may be obtained in a manner similar to least-squares 
fitting from the curvature matrix 
$\partial^2(\alpha S -\chi^2/2)\partial f_i\partial f_j$.

As was mentioned above,
the SDI interaction exhibits a bad isospin
dependence that can be corrected by adding the term $a{\hat T}^2$ to
the Hamiltonian. Unfortunately, this additional term has a bad sign
in the MCSM and cannot be evaluated directly. The extrapolation
method of Ref.~\cite{r:r11} could be used, but at a significant 
computational cost and larger statistical errors.
Instead,
for even-even nuclei, where the low-lying states are unaffected by the
additional term, a perturbative approach may be more useful. Towards this
end, $a{\hat T}^2$ is added to the SDI Hamiltonian, and the energy
is evaluated by expanding the $a{\hat T}^2$
terms in the exponential in Eq.~(\ref{e:energy}) to first order in 
$\beta$. Unfortunately, a further limitation
is imposed due to computational limitations that make it  
impractical to evaluate the expectation value of $n$-body operators 
in the MCSM beyond 
$n=2$. Given these considerations, the first-order correction to the 
energy is estimated as 
\begin{equation}
E_{corr}=\beta a\frac{\partial\langle {\hat T}^2\rangle}{\partial \beta}
+a\langle {\hat T}^2\rangle.
\label{e:tcorr}
\end{equation}
In comparison with exact results, Eq.~(\ref{e:tcorr}) works quite
well, although it tends to ``over correct'' by approximately 10\%.
This over correction, which may be due to the neglected
$({\hat T}^2)^2$ terms in the expansion, can be damped
by multiplying $E_{corr}$ by the factor 
$e^{-\beta a\langle {\hat T}^2\rangle}$. For illustrative purposes,
both $E(\beta)$ and $\ln Z(\beta)$ are shown in Fig.~\ref{fig:fig2} 
for $^{20}$Ne 
using the SDI interaction with and without the 
iosospin correction factor. In the figure, the solid and dashed lines 
represent $E(\beta)$ and $Z(\beta)$ obtained 
by using the exact eigenvalues of a shell-model diagonalization for 
the SDI and ${\rm SDI}+a{\hat T}^2$ Hamiltonians,
respectively. The dotted line represents the results obtained by
adding $E_{corr}$ to the SDI results. From the figure, it is seen that
the corrected values very accurately reproduce both the
energy and partition function of the ${\rm SDI}+a{\hat T}^2$ Hamiltonian.

Shown in Fig.~\ref{fig:fig3} 
are the reconstructed level densities obtained for $^{24}$Mg
using the ${\rm SDI}$ (top) and ${\rm SDI}+1.507{\hat T}^2$ (bottom) 
Hamiltonians, respectively. For the reconstruction, $E(\beta)$ was normalized
relative to the ground state value, which was evaluated to be 
$-76.844(200)$~MeV by performing MCSM calculations for 
$1.75\le\beta\le 2.25$~MeV$^{-1}$ and extrapolating for 
$\beta\rightarrow\infty$. The histogram in both panels represents the
exact level density for $^{24}$Mg using the USD Hamiltonian, and was
obtained by direct diagonalization and placing the 28503 ($J_z=0$) 
eigenvalues 
into 1~MeV bins while including the $2J+1$ degeneracy. Because of the
near Gaussian structure of the level density, 
$\rho(E)$ is only plotted up to 45~MeV, which is slightly larger than the
centroid of the USD level density ($\sim 42$~MeV). From the top panel, it is 
clear that the level density obtained from the SDI Hamiltonian alone 
is much too compressed (centroid at $\sim 36$~MeV), and 
over predicts the $\rho(E)$ by nearly a factor of two 
for $E\le 20$~MeV. On the other hand, the corrected 
level density represents a considerable improvement (centroid $\sim 40$~MeV), 
and gives a reasonable representation of the USD level density for 
excitation energies $\le 20$~MeV.

At this point, it is important to note an important limitation in using the
shell model to estimate the nuclear level density. Because of the fact that
all calculations are by necessity limited to a finite model space, the
shell-model level density will always have a Gaussian character, and at some
point will always underestimate the true level density because of the presence
of states representing excitations outside of the model space.
For the most part, the shell model is best suited to describe states at lower
excitation energies. Hence, it is not realistic to expect that  
the shell model could provide an estimate of the level density
for excitation energies around 40-50~MeV. In the case of a single major
oscillator shell, such as the {\it sd}-shell used here, the results are
comparable to experimental data for $E\le 10$~MeV.
On the other hand, by enlarging the model space to 
include more configurations, say another major oscillator shell such as the 
{\it fp}-shell, the method presented here can be used to make 
estimates for the level density
up to excitation energies of the order 15-20~MeV.
In this case, however, it will also be necessary to account for 
states of differing parity, as well as spurious excitations of the center
of mass.

To conclude, the procedure outlined here can successfully describe the
level density of a realistic nuclear system, and 
plans are currently underway to compute level densities for 
$^{22}$Mg, $^{26}$Si, $^{30}$S, and $^{34}$Ar, which
are needed to make estimates of $(\alpha,\gamma)$ reaction rates of 
astrophysical interest~\cite{r:r15}. 
In addition, the method will also be applied to
$^{162}$Dy and $^{172}$Yb where experimental data~\cite{r:r16} for $\rho(E)$ 
exists for excitation energies up to the neutron separation energy. In
these nuclei, since the protons and neutrons occupy different major shells,
the problems associated with the higher isospin states pointed out here
are most likely to be mitigated.

Discussions with C.~W.~Johnson and J.~P.~Draayer are gratefully
acknowledged. This work was supported in part by NSF Cooperative
agreement No. EPS~9550481, NSF Grant No. 9603006, and DOE contract
DE--FG02--96ER40985. 

\newpage
\bibliographystyle{try}

\newpage
\begin{table}
\caption{Excitation energies (in MeV) for the first $J=0$ states of each
isospin for the SDI and USD interactions for $^{24}$Mg and $^{32}$S. 
The number of states for
each isospin contributing to the total level density is also tabulated.}
\begin{tabular} {crrrrr}
 & \multicolumn{2}{c}{$^{24}$Mg} & \multicolumn{2}{c}{$^{32}$S} & \\
$T$ & USD & SDI & USD & SDI & \# States \\
\tableline
0  &  0.000 &  0.000 &  0.000 &  0.000 & 166,320 \\
1  & 12.872 &  9.351 &  7.312 &  4.735 & 332,640 \\
2  & 15.425 &  7.903 & 12.060 &  6.434 & 237,600 \\
3  & 33.923 & 22.528 & 33.468 & 22.268 &  83,160 \\
4  & 46.040 & 26.251 & 45.404 & 26.328 &  11,880 \\
\end{tabular}
\label{tab:tab1}
\end{table}

\begin{table}
\caption{Comparison between the USD and SDI excitation energies (in MeV) 
for the lowest few $T=0$ states in $^{24}$Mg. Shown are the first ten states
as predicted by the USD interaction, and then the lowest $J=7$ and
8 states.}
\begin{tabular} {cccccc}
$J$ &  USD & SDI & $J$ &  USD & SDI  \\
\tableline
0  &  0.000 &  0.000 & 0 &  7.561 &  6.778 \\
2  &  1.509 &  1.973 & 1 &  7.764 &  7.253 \\
2  &  4.122 &  4.146 & 5 &  7.883 &  8.281 \\
4  &  4.378 &  4.845 & 6 &  8.263 &  8.460 \\
3  &  5.097 &  5.326 & 7 & 12.283 & 12.596 \\
4  &  5.934 &  5.314 & 8 & 12.088 & 10.974 \\
\end{tabular}
\label{tab:tab2}
\end{table}
\newpage

\begin{figure}
\caption{Comparison of binding energies for oxygen isotopes as a function 
of neutron number obtained with the
USD (solid line) and ${\rm SDI}+0.546{\hat T}^2$ (dotted line) 
Hamiltonians, respectively.}
\label{fig:fig1}
\end{figure}

\begin{figure}
\caption{Plot of the energy, $E(\beta)$, and partition function,
$Z(\beta)$ for $^{20}$Ne, as a function of $\beta$ for the SDI and 
${\rm SDI}+a{\hat T}^2$ Hamiltonians. The solid and dashed lines represent
the SDI and ${\rm SDI}+a{\hat T}^2$ 
results obtained from the exact eigenvalues, while the dotted line
shows the SDI results corrected via Eq.~(\ref{e:tcorr}).}
\label{fig:fig2}
\end{figure}

\begin{figure}
\caption{Reconstructed level densities obtained for $^{24}$Mg using the
SDI (top panel) and the ${\rm SDI}+1.507{\hat T}^2$ (bottom panel) 
Hamiltonians, respectively. The histogram in both panels represents the
$^{24}$Mg level density obtained with the USD interaction.}
\label{fig:fig3}
\end{figure}

\
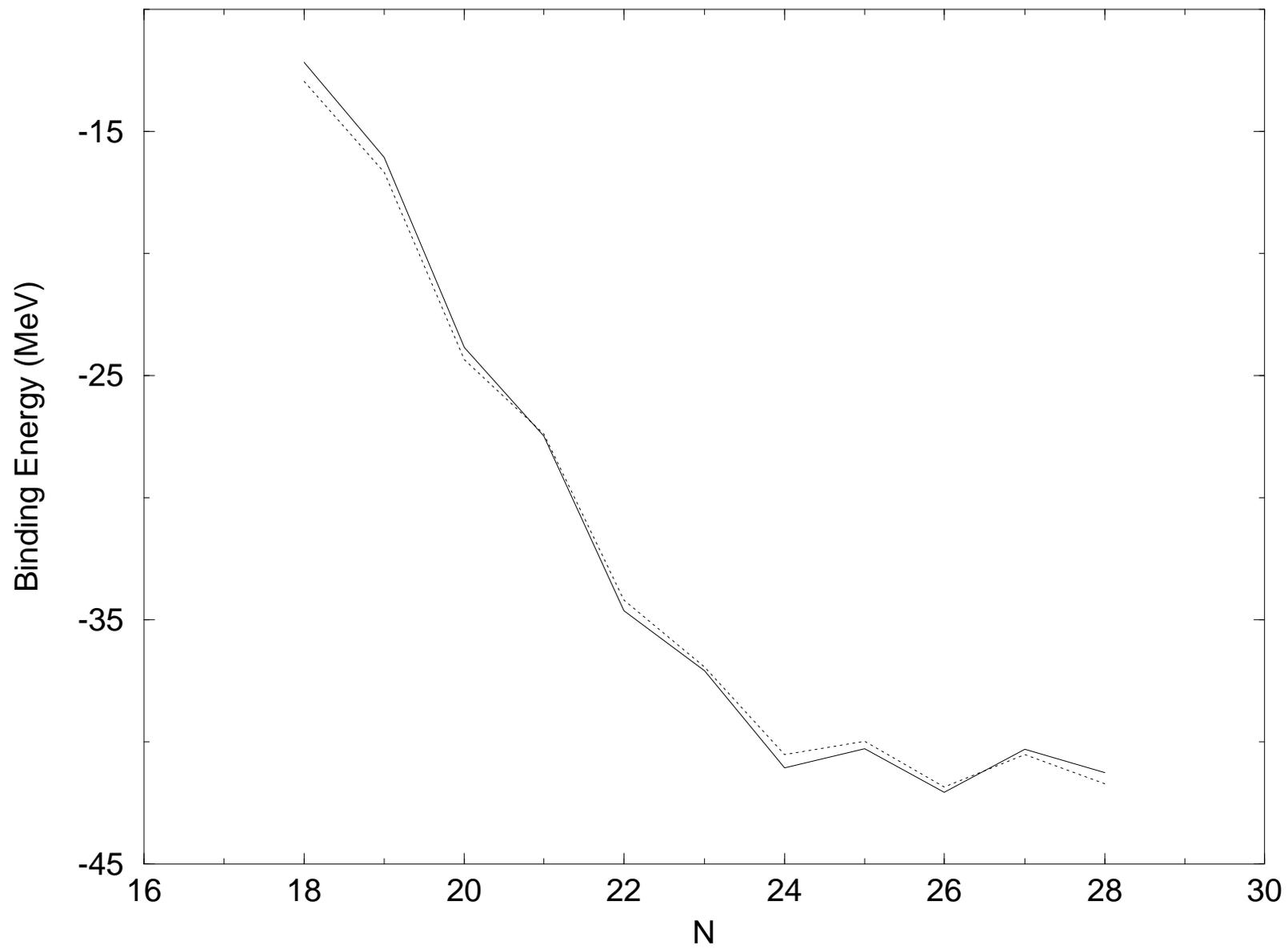
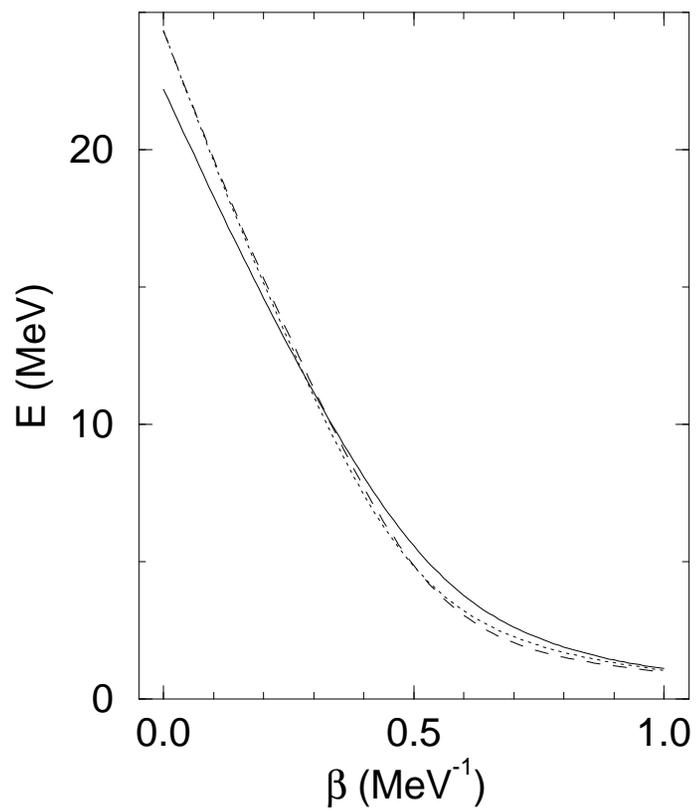
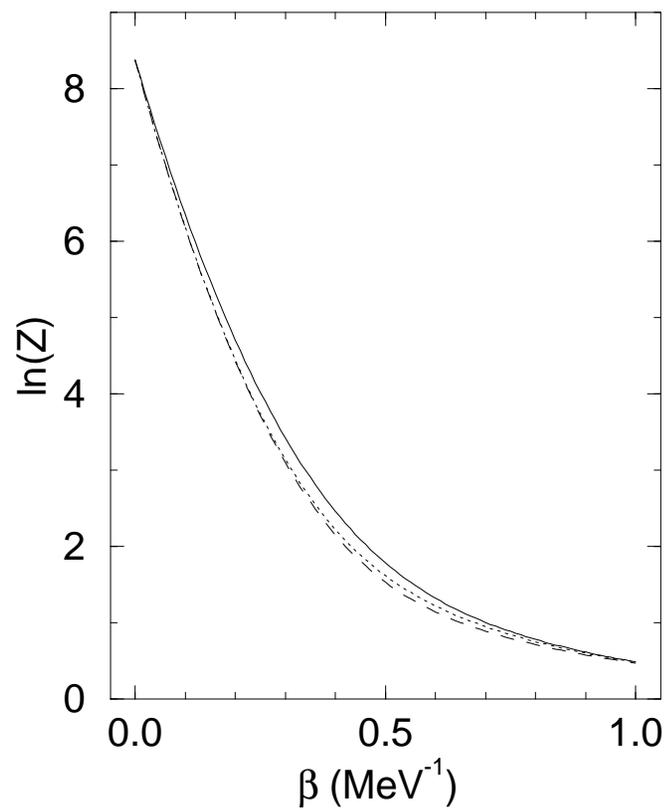
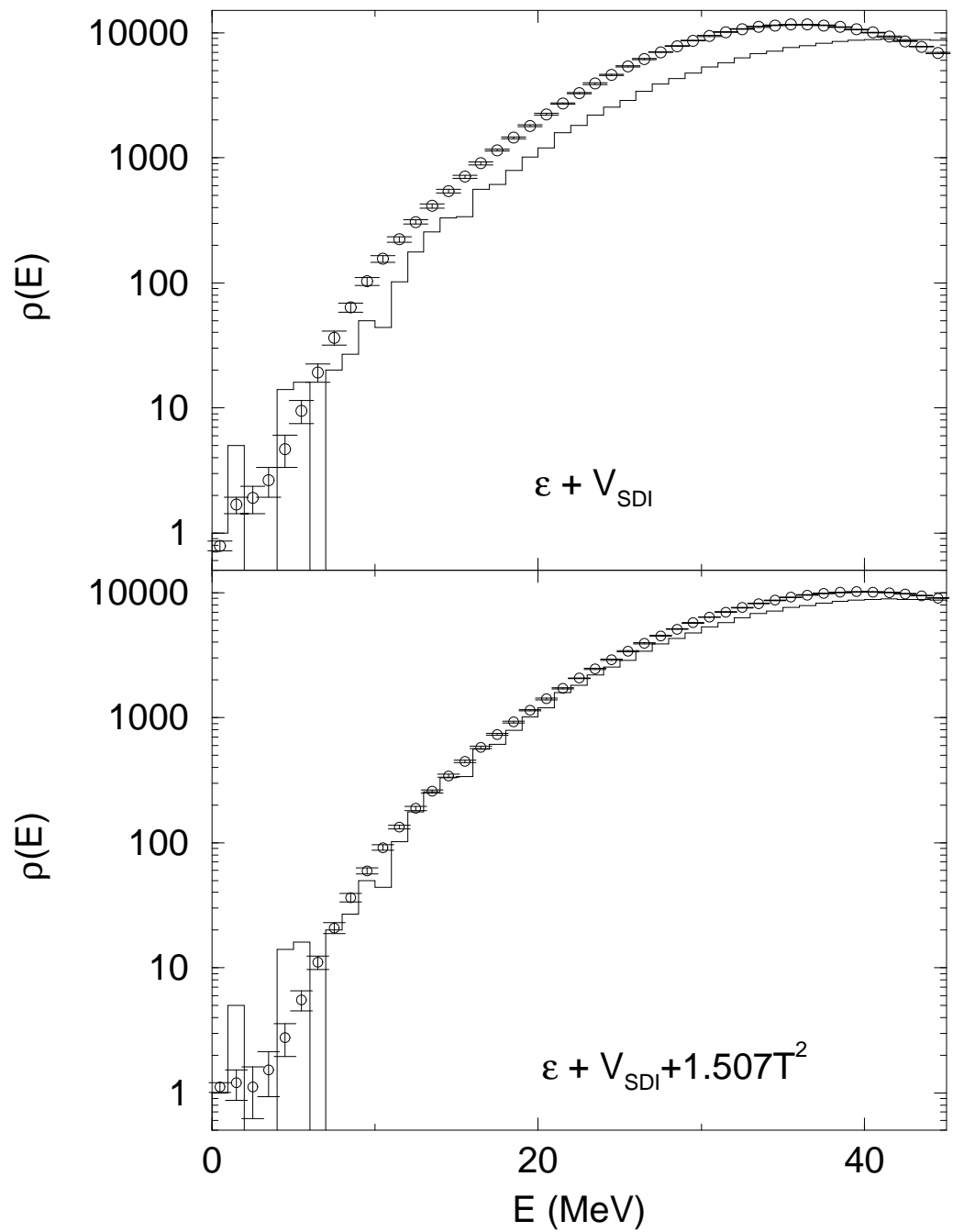
\end{document}